\documentclass{statsoc}
\usepackage[a4paper, top=1.1in, bottom=1in, left=1in, right=1in]{geometry}
\usepackage{graphicx} 
\usepackage{amsmath}
\usepackage{amssymb}
\usepackage[latin1]{inputenc}
\usepackage[table]{xcolor}
\usepackage[ruled, linesnumbered]{algorithm2e}
\usepackage{multirow, makecell} 
\usepackage{booktabs}
\usepackage{natbib}
\usepackage[sans]{dsfont}
\usepackage{enumitem}
\usepackage{lineno}
\usepackage{hyperref}
\usepackage{xcolor,color}
\usepackage{footnote}
\usepackage{etoolbox}

\makeatletter
\patchcmd{\@makecaption}
  {\parbox}
  {\advance\@tempdima-\fontdimen2} 
  {}{}
\makeatother  
\makesavenoteenv{tabular}
\makesavenoteenv{table}

\definecolor{lightgray}{gray}{0.9}

\SetCommentSty{mycommfont}
\SetKwComment{Comment}{$\triangleright$\ }{}

\newcommand{\Indicator}{\operatorname{\mbox{\large $\mathds{1}$}}}
\newcommand{\ex}{\mathbb{E}}

\newcommand{\upmydate}{19} 

\newcommand{\wtot}{121,464,666\ } 
\newcommand{\wdeath}{2,684,093\ } 
\newcommand{\ptot}{238\ } 

\newcommand{\utot}{29,317,562\ } 
\newcommand{\udeath}{532,971\ } 

\newcommand{\btot}{11,693,838\ } 
\newcommand{\bdeath}{284,775\ } 

\newcommand{\itot}{11,514,331\ } 
\newcommand{\ideath}{159,370\ } 

\newcommand{\mtott}{1,734,503\ } 
\newcommand{\mdeath}{182,009\ } 

\newcommand{\mtotbase}{5,839,528\ } 
\newcommand{\mtot}{1,584,288\ } 

\newcommand{\mact}{585,034\ } 
\newcommand{\mtottrain}{1,484,288\ } 

\definecolor{goldenrod4}{rgb}{0.55,0.41,0.08}
\definecolor{darkgoldenrod1}{rgb}{1,0.73,0.06}
\definecolor{darkorange1}{rgb}{1,0.5,0}
\definecolor{orangered1}{rgb}{1,0.27,0}
\definecolor{red3}{rgb}{0.8,0,0}

\title[COVID-19 and mortality]{COVID-19 Clinical footprint to infer about mortality} 

\author{Carlos E. Rodr\'iguez\thanks{Corresponding author. Department of Probability and Statistics, Universidad Nacional Aut\'onoma de M\'exico, M\'exico. e-mail: carloserwin@sigma.iimas.unam.mx}}
\address{Universidad Nacional Aut\'onoma de M\'exico, M\'exico.}

\author[Carlos E. Rodr\'iguez \& Rams\'es H. Mena]{Rams\'es H. Mena}
\address{Universidad Nacional Aut\'onoma de M\'exico, M\'exico.}

\begin{document}
\begin{abstract}  
Information of 1.6 million patients identified as SARS-CoV-2 positive in Mexico is used to understand the relationship between comorbidities, symptoms, hospitalizations and deaths due to the COVID-19 disease. Using the presence or absence of these latter variables a clinical footprint for each patient is created. The risk, expected mortality and the prediction of death outcomes, among other relevant quantities, are obtained and analyzed by means of a multivariate Bernoulli distribution. The  proposal considers all possible footprint combinations resulting in a robust model suitable for Bayesian inference. 
\end{abstract}

\section{Introduction}
On 31 December 2019, the World Health Organization (WHO) received a troubling report from Chinese health officials, \cite{who}. A mystery pneumonia had sickened dozens of people in Wuhan, the capital of Hubei Province in China. A virus, that we know now as SARS-CoV-2, had been transmitted from an unknown animal host to humans and since then has turned up lives worldwide with an unprecedented speed. 

As of \upmydate$^{th}$ March, 2021 the WHO database has confirmed \wtot cases SARS-CoV-2 virus globally with \wdeath reported deaths caused by   COVID-19 disease from \ptot countries. The most affected countries are the United States with \udeath deaths out of \utot confirmed cases; Brazil with \bdeath deaths out of \btot cases; Mexico with \mdeath deaths out of \mtott confirmed cases and  India with \ideath deaths out of \itot confirmed cases, \cite{who}. 


\subsection*{Clinical characteristics of COVID-19}
The incubation period of COVID-19, defined  as the time between exposure to the virus and symptom onset, is on average five to six days, but it can be as long as 14 days, \cite{incub}. The symptoms of COVID-19 range from those that might not be noticeable to severe life-threatening illness.  Some infected people have no symptoms, known as asymptomatic or pre-symptomatic carriers, \cite{assympt}.

According to the WHO most infected people will develop mild to moderate illness and recover without hospitalization. The WHO divides COVID-19 symptoms in three groups: {\it most common symptoms}; fever, dry cough and tiredness, {\it less common symptoms}; aches and pains, sore throat, diarrhea, conjunctivitis, headache, loss of taste or smell, a rash on skin, or discoloration of fingers or toes, and {\it serious symptoms}; difficulty breathing or shortness of breath, chest pain or pressure, loss of speech or movement, \cite{who}. 

It is important to mention that individual symptoms appear to have poor diagnostic properties. Indeed, based on currently available data, neither absence nor presence of any symptoms are accurate enough to rule in or rule out the disease, \cite{Struyf20}. Thus, the gold standard for COVID-19 diagnosis is the laboratory technique known as RT-PCR test (Reverse-Transcription Polymerase Chain Reaction test), however there are other alternatives, see \cite{OLIVEIRA2020}.
 
 \subsection*{Comorbidities and its effects in COVID-19 patients}
People of any age who have underlying medical conditions, such as hypertension and diabetes, have shown worse prognosis, \cite{comorb}. Diabetic patients have increased morbidity and mortality rates and have been linked to more hospitalization and intensive care unit (ICU) admissions, \cite{SINGH2020303}. People with chronic obstructive pulmonary disease (COPD) or any respiratory illnesses are also at higher risk for severe illness from COVID-19,  \cite{COPD}.

\subsection*{Impact of sex and age on COVID-19 outcomes}
From the first reports from China a sex imbalance with regard to the fatality rate of COVID-19 patients has been detected. Case fatality rates reported in China, Italy, Spain, France, Germany, and Switzerland support the view that a consistent biological phenomenon is operating, accounting for a higher case fatality in men. Such observation  is independent of country specific demographics and testing strategies, \cite{sex}.

Age has also been identified as a variable with high impact over the mortality rate of COVID-19 cases. All age groups appear to have significantly higher mortality compared with the immediately younger age group, \cite{age}.

\subsection*{Objectives and methods}
The aim of this research is to use the available data from the COVID-19 pandemic in Mexico to gain insight into the COVID-19 disease. The first particular objective is to understand the relationship between comorbidities, symptoms, mortality risk and hospitalization. The second is to identify differences by sex and age group. And the third is mortality prediction, that is given the commodities, symptoms, age and sex of a patient identified as SARS-CoV-2 positive, then estimate the mortality risk of a given patient. It is important to observe that almost all relevant variables in this study are binary and indicate the presence or absence of a symptom/comorbidity or whether a patient has died, has been hospitalized, or not.

A straightforward first strategy to analyze the impact of comorbidities and symptoms over death outcomes and hospitalizations under a binary data setting is to use contingency tables. Hence, basic probabilities such as $P(Y_j = 1)$ and $P(Y_i=1|Y_j = 1)$ can be easily computed via the observed frequencies. This can be extended to account for higher order interactions of the form $P(Y_i  |Y_j, Y_r )$. The idea to compute these probabilities efficiently is to concatenate the observed combinations $y_jy_r$ and then to obtain the correspondent frequencies. The same ideas could be applied to obtain probabilities such as $P(Y_{i}  |Y_{j_1}, Y_{j_2},  \ldots, Y_{j_k})$. Moving to a modelling framework, a second vanilla strategy is to use logistic regression models. Under such a strategy, an idea would be to estimate death outcome probabilities using all available variables. The challenge under this setting is to select a small subset of variables that could describe the death outcomes effectively. However, it is not possible to obtain all the summaries of interest as the joint distribution over all variables is sometimes needed and not robustly available for such regression method. A third and possibly more general approach is to use a multivariate binary distribution.

To achieve the aforementioned objectives via the generation of relevant information summaries such as those obtained via contingency tables while maintaining the predictive capabilities of the logistic regression, we have chosen the multivariate Bernoulli distribution \citep[see][]{dai2013}. In this case all the relevant quantities are obtained in closed form and its implementation is straightforward. This model is a generalization of the well known Bernoulli distribution that takes the value $1$ (success) with probability $p$ and the value $0$ (failure) with probability $1-p$. In the multivariate case each observation is a vector of, let us say, $k$ successes/failures. The multivariate Bernoulli distribution assigns positive probability to each of the $2^k$ possible combination of successes/failures. It is important to note that this model can estimate not only the main effects and the interactions between pairs of variables, but is also capable of modeling higher-order interactions. For us the information of each patient is represented by a $k$ vector of ones and zeros: $1$ (presence) or $0$ (absence) of each comorbidity, symptom, hospitalization and death. Such data composition will be referred to as the COVID-19 footprint for each patient. With the multivariate Bernoulli distribution and the footprint for each patient is relatively simple to use a Bayesian approach to obtain meaningful inference. Furthermore, under such an approach it is possible to obtain inferences about quantities such as the mortality risk given certain comorbidity, sex and age group, the mortality risk given certain symptom, sex and age group. Indeed, this will be done for all patients identified as SARS-CoV-2 positive by a laboratory test.

In Section~\ref{sec:2} information about the prevalence of the comorbidities that the Mexican population suffers the most are provided, and the data used for this study is described. In Section~\ref{sec:3}, the multivariate Bernoulli model is introduced and some of its properties discussed. Section~\ref{sec:4}, deals with the Bayesian ideas to make inference via the multivariate Bernoulli distribution. In Section~\ref{sec:5}, the available data from the pandemic in Mexico is analyzed and some interesting results are presented and described. Finally, a discussion is provided in Section~\ref{sec:7}.
 
 \section{COVID-19 pandemic in Mexico}\label{sec:2}
In Mexico, the first cases of SARS-CoV-2 were detected on the $29^{th}$ of February 2020 and to the date of writing (\upmydate$^{th}$ March, 2021), this represents \mtott confirmed cases and \mdeath deaths. With this numbers Mexico has become the country with the third highest death toll with coronavirus, with only the US and Brazil recording greater numbers. 

\subsection*{Diabetes, obesity and hypertension in Mexico}
The World Health Organization has said people with underlying medical problems like high blood pressure, heart and lung problems, diabetes, or cancer, are among those most vulnerable to severe cases of the new coronavirus disease, along with the elderly \cite{who}. 

Over the past 30 years, Mexico has become one of the countries in the world most heavily affected by the global epidemic of obesity. It is now the second country worldwide with obesity prevalence. Between 2006, 2012 and 2018, overweight or obesity (both) prevalence increased from 69.5\% to 71.3\%  and then to 75.2\% (respectively) in population of 20 years and over, while the rate of obesity alone rose from 30\% in 2006 to 32.4\% in 2012 and then to 36.1\% in 2018 (estimates). Also, Mexico is now one of the countries with the highest child obesity rates in the world with one in three children being overweight or obese. Diabetes, the chronic disease most directly linked with obesity, is spreading rapidly and in Mexico in 2018 affected 10.3\% of the adult population (aged over 20 years old), while in 2012 it affected 9.2\%. 

High blood pressure, or hypertension, has less noticeable symptoms. But if untreated, it increases the risk of serious problems such as heart attacks and strokes. In Mexico, the prevalence of hypertension in 2012 was of 30.2\% and in 2018 of 32.7\%. However, these estimates may not be comparable as there was a methodological change in the National Health and Nutrition Survey\footnote[2]{\url{https://www.inegi.org.mx/programas/ensanut/2018/}}(NHNS) of 2018. Digital baumanometers were introduced; being less susceptible to measurement error, they provide interviewers with better estimates of blood pressure values, see \cite{hyper2}.

All the information in this section has been obtained via the NHNS. This survey is conducted every six years by the National Institute of Statistics and Geography which is an autonomous agency of the Mexican Government dedicated to coordinate the National System of Statistical and Geographical Information of Mexico.

\subsection*{Data and variables}
The source of information for this work is the database of the National Epidemiological Surveillance System for monitoring possible cases of COVID-19 in Mexico (SINAVE/SISVER for its acronym in Spanish), coordinated by the Secretary of Health (Spanish: Secretar\'ia de Salud \footnotemark[1]{Is the government agency in charge of all social health services in Mexico, a very important  part of the Mexican health system.}). The SINAVE/SISVER platform considers cases that are suspected of COVID-19. People who have had flu like symptoms, or that believe to have been infected with the SARS-CoV-2 virus, are entitled to attend any public or private health service in Mexico, and after an initial examination is suspected to suffer from the COVID-19 disease are registered on this database. 

By March $19^{th}$, 2021 this database had information for a total of \mtotbase suspected cases and 115 variables. This is the latest update of this data base that we have. For the relevance of this analysis we will use only the following variables:
\begin{itemize}
\item Result from the lab test (positive or negative for the presence of the SARS-CoV-2 virus in a blood test). We work with the information of \mtot positive cases.
\item Gender (Female and Male) and age group (four groups $< 20$, $[20, 40)$, $[40, 60)$ and $>60$).
\item Symptoms: fever (1), cough (2), ears pain (3), difficulty breathing (4), irritability (5), diarrhea (6), chest pain (7), chills (8), headache (9), muscle pain (10), joint pain (11), attack general state (12), nasal discharge (13), increased respiratory frequency and depth (14), vomiting (15), abdominal pain (16), conjunctivitis (17), blue color lack of oxygen (18) and sudden onset of symptoms (19).
\item Comorbidities: chronic kidney failure (1), COPD (2), heart disease (3), diabetes (4), immunosuppression (5), hypertension (6), obesity (7), smoking (8) and asthma (9).
\item Disease outcomes: hospitalization (1) and death (2).
\end{itemize}
Symptoms, comorbidities, sex and outcomes of the disease are dichotomous variables, namely each patient has only two possible values presence ($1$) or absence ($0$) of the symptom and/or comorbidity. Sex is also taken as dichotomous, male (1) and female (0). Also the variables recording if the patient has been hospitalized (1) or not (0) or the patient has died (1) or has not (0) are decoded in a similar manner. To include the age groups; $< 20$, $[20, 40)$, $[40, 60)$ and $>60$, under the framework of the multivariate Bernoulli distribution described below, these can be treated as $(1,0,0,0)$, $(0,1,0,0)$, $(0,0,1,0)$ and $(0,0,0,1)$ respectively. 

Therefore, in this study the clinical footprint of each patient that has been identified as SARS-CoV-2 positive is given by
\begin{eqnarray*}
\boldsymbol{y}_{35}&=&(\mbox{sex}, \mbox{age group}_{1:4}, \mbox{comorbidity}_{1:9}, \mbox{symptom}_{1:19}, \mbox{hospitalized}, \mbox{death}),\\
&=& (s, g_{1}, \ldots, g_{4}, j_{1}, \ldots, j_{9}, l_{1},\ldots, l_{19}, h, d),
\end{eqnarray*}
which is a vector of zeros and ones of length 35. In other words, our dataset is a matrix of zeros and ones of dimensions $\mtot \times 35$.

\subsubsection*{Open access data}
The Secretary of Health via the General Directorate of Epidemiology (Spanish: Direcci\'on General de Epidemiolog\'ia)  updates daily the data base with the suspected cases and 48 variables, these data can be accessed using the following link \url{https://www.gob.mx/salud/documentos/datos-abiertos-152127}. Additional variables are available upon request via the platform \url{http://covid-19.iimas.unam.mx/}, as described in \cite{pplat}.

\subsubsection*{ COVID-19 case}
Health authorities in Mexico classify a patient as a COVID-19 case if he/she fells in one of the following three categories:
\begin{itemize}
\item Confirmed by epidemiological clinical association. Confirmed by association applies when the case had contact with a COVID-19 case, and the latter is registered in SISVER plataform. The case was not tested or the test was invalid.
\item Confirmed by dictamination committee.  Confirmed by ruling out only applies to deaths when the case was not tested or a test was taken, but it was invalid.
\item Confirmed by SARS-CoV-2.  The case has a laboratory test or antigenic test and was positive for SARS-CoV-2, regardless of whether the case has a clinical epidemiological association.
\end{itemize}
As of $19^{th}$ of March 2021 there have been $1,734,503$ confirmed cases: $120,944$ by association, $5,307$ by dictamination and $1,608,252$ by a laboratory test. In this work we only consider cases that have tested positive for SARS-CoV-2 in a laboratory test.

\subsubsection*{Missing data}
The complete data base has \mtotbase records. Focusing only in the positive cases by laboratory test, there are 1,608,252 patients. At this instance, there are no missing values in the variables with information about deaths, hospitalizations, sex and age. However, if symptoms (19) and comorbidities (9) are included, there are $0.24\%$ of cells, in the $1,608,252 \times 28$  database, with missing values. The percentage of missing values over the commorbidities ranges from $0.25\%$ (obesity) to $0.29\%$ (diabetes), while for the symptoms ranges from $0.08\%$ (difficulty breathing) to $0.5\%$ (sudden onset of symptoms). Assuming a missing completely at random mechanism, we followed a case deletion strategy to obtain the \mtot registers that we work in this analysis.

\subsubsection*{Uncertainty and bias}
As an additional source of uncertainty we have that the information about the symptoms and comorbidities is mainly self declared by patients and the Mexican health authorities know there must be many patients not fully aware if they suffer from certain comorbidities. 

The number of SARS-CoV-2 positive tests has turned into an important indicator, it has been used to decide whether or not nations or regions around the world can open their economies. This is assessed using what has been called the percent positive which is simply the percentage of all coronavirus tests performed that are actually positive. The WHO recommended that the percent positive remain below 5\% for at least two weeks before governments consider reopening after a lock down period, \cite{who}. In Mexico, the percent positive since March 2020 has been larger than 30\% and there even have been periods where it almost reached $50\%$. These very high percentages are easy to explain, the Mexican health authorities operate with limited resources, which forbids a widespread testing. Thus mainly patients with COVID-19 symptoms have been tested. Hence, it is clear that there is bias in the data base of \mtotbase suspected cases, and is far from a random sample. 

All this said, the \mtot positive cases are expected to provide us with important information for inference.

\section{The model}\label{sec:3}
Let $\boldsymbol{Y}_k = (Y_1, Y_2, \ldots, Y_k)$ be a $k$-dimensional random vector of possibly correlated Bernoulli random variables and let $\boldsymbol{y}_k=(y_1, y_2, \ldots, y_k)$ be a realization of  $\boldsymbol{Y}_k$. In our case $k=35$.  The multivariate Bernoulli distribution can be described via its mass probability function  
\begin{eqnarray*}
p(\boldsymbol{Y}_k= \boldsymbol{y}_k|\boldsymbol{w}_{2^k}) &=& w_k(0, 0, 0, \ldots, 0)^{\{\prod_{j=1}^k (1-y_j)\}}\\
 &\times& w_k(1, 0, 0, \ldots, 0)^{\{y_1 \prod_{j\neq 1} (1-y_j)\}}\\
 &\times& w_k(0, 1, 0, \ldots, 0)^{\{y_2 \prod_{j\neq 2} (1-y_j)\}}\\
 &\times& \ldots \qquad \ldots \qquad \ldots \\
 &\times& w_k(1, 1, 0, \ldots, 0)^{\{y_1 y_2 \prod_{j \neq 1, 2} (1-y_j)\}}\\
  &\times& \ldots \qquad \ldots \qquad \ldots \\
 &\times& w_k(1, 1, 1, \ldots, 1 ) ^{\{\prod_{j= 1}^k y_j\}}.
\end{eqnarray*}
where 
\begin{equation}
\boldsymbol{w}_{2^k} =  (w_k(1, 0, 0, \ldots, 0),  w_k(0, 1, 0, \ldots, 0), \ldots,  w_k(1, 1, 1, \ldots, 1))\label{wfull}
\end{equation} 
is the vector of probabilities associated to each possible outcome, and there are $2^k$ possible outcomes. 

Taking $\boldsymbol{\rho}^k = \{\{{\rho^k_{l,j}\}_{l=1}^{2^k}}\}_{j=1}^k$ as the matrix where the rows are the $2^k$ possible outcomes of the random vector $\boldsymbol{Y}_k$ and  $\boldsymbol{\rho}^k_{l,.}$ as the $k$ dimensional vector of the $l$ possible outcome. It is possible to write the above mass probability function in  shorter forms, that is 
\begin{eqnarray}
p(\boldsymbol{Y}_k= \boldsymbol{y}_k|\boldsymbol{w}_{2^k}) &=& \sum_{l=1}^{2^k}w_k(\boldsymbol{\rho}^k_{l,.})\ \gamma_l(\boldsymbol{y}_k), \nonumber \\
&=& \prod_{l=1}^{2^k} w_k(\boldsymbol{\rho}^k_{l,.})^{\gamma_l(\boldsymbol{y}_k)}.  \label{eq1}
\end{eqnarray}
where $\gamma_l(\boldsymbol{y}_k) = \Indicator(\boldsymbol{\rho}^k_{l,.} = \boldsymbol{y}_k)$. 

These expressions clearly indicate that we have  $2^k$ parameters, namely one parameter value per each possible footprint. Indeed, much in the same spirit as for the univariate Bernoulli distribution, the above multivariate version constitutes a dense model in its support. Namely all possible combinations are considered. The above finite mixture representation suggests  the possibility of using a discrete random probability measure as a potential model, namely a Bayesian nonparametric distribution. However, such an approach, e.g. via the Dirichlet processes, would require and infinite number of footprints and a diffuse baseline distribution to ease its implementation. While potentially possible some adaptations and interpretations would be at hand, so we prefer to keep it simple.


\subsection*{Properties}
The marginal distribution of the random vector $\boldsymbol{Y}_k$ which follows a multivariate Bernoulli distribution with density function (\ref{eq1}) to any order is still a multivariate Bernoulli distribution. Thus, for $s < k$
\begin{eqnarray*}
p(\boldsymbol{Y}_s = \boldsymbol{y}_s | \boldsymbol{w}_{2^s})  &=& \sum_{y_{s+1} = 0}^1\cdots \sum_{y_k = 0}^1 p(\boldsymbol{Y}_k= \boldsymbol{y}_k|\boldsymbol{w}_{2^k}),\\
&=&  \prod_{r =1}^{2^{s}} w_{s}(\boldsymbol{\rho}^{s}_{r,.})^{\gamma_l(\boldsymbol{y}_s)  }, 
\end{eqnarray*}
is a multivariate Bernoulli distribution where 
$$w_{s}(\boldsymbol{\rho}^{s}_{r,.})=  \sum_{l=1}^{2^k} \left[\Indicator{(\boldsymbol{\rho}^k_{l,1:s} = \boldsymbol{\rho}^{s}_{r,.} )}\ w_k(\boldsymbol{\rho}^k_{l,.}) \right],\ \mbox{for}\ r=1,\ldots, 2^{s}, $$
with $\boldsymbol{\rho}^k_{l,1:s}$ as the first $s$ observations of the  $l$ outcome of the random variable $\boldsymbol{Y}_k$, and $\boldsymbol{\rho}^s_{r,.}$ as the $s$ dimensional vector of the $r$ possible outcome of the marginalized random variable $\boldsymbol{Y}_s$.  

As a straightforward consequence, for $j \in \{1, \ldots, k\}$, we have 
 \begin{equation}
 p(Y_j = y_j |w_j) = w_j ^{y_j} (1-w_j)^{1-y_j},\label{py}
 \end{equation}
 where  
 \begin{equation}
 w_j = \sum_{l=1}^{2^k}  \Indicator{(\rho^k_{l,j} =1)}\ w_k(\boldsymbol{\rho}^k_{l,.}).\label{wj}
 \end{equation}
Thus each $Y_j$ follows a Bernoulli distribution as expected. Also, the conditional distribution of $\boldsymbol{Y}_s$ given the rest is also multivariate Bernoulli distribution ($s < k$).  

Finally, in this brief outline, the covariance between any two random variables $Y_j$ and $Y_s$, is given by 
\begin{eqnarray*} 
\mbox{Cov}(Y_j, Y_s) &=& p(Y_j = 1, Y_s = 1|\boldsymbol{w}_{4}) -  w_j w_s, \nonumber \\
&=& w_{j,s} - w_j w_s, 
\end{eqnarray*}
where 
\begin{equation}
w_{j,s} = \sum_{l=1}^{2^k}  \Indicator{(\rho^k_{l,j} =1)}  \Indicator{(\rho^k_{l,s} =1)}\ w_k(\boldsymbol{\rho}^k_{l,.}). \label{wjs}
\end{equation}

Then, the correlation is given by 
\begin{eqnarray} 
\mbox{Corr}(Y_j, Y_s) = \frac{w_{j,s} - w_j w_s}{\sqrt{w_j(1-w_j) }\sqrt{w_s (1-w_s)}}. \label{eq3}
\end{eqnarray}

\subsection{Bayesian inference}\label{inference}
For a random sample $\boldsymbol{y}_{1,k}, \boldsymbol{y}_{2,k}, \ldots, \boldsymbol{y}_{n,k}$ the likelihood of the multivariate Bernoulli distribution is given by 
\begin{eqnarray*}
\prod_{i=1}^n p(\boldsymbol{Y}_k= \boldsymbol{y}_{i, k}|\boldsymbol{w}_{2^k}) &=& \prod_{i=1}^n \left[\prod_{l=1}^{2^k} w(\boldsymbol{\rho}^k_{l,.})^{\gamma_{l,i}(\boldsymbol{y}_{i,k}) } \right]\\
&=& \prod_{l=1}^{2^k} w(\boldsymbol{\rho}^k_{l,.})^{ r_l },
\end{eqnarray*}
where 

\begin{equation}
r_l = \sum_{i=1}^n  \gamma_{i, l}(\boldsymbol{y}_{i,k}) = \sum_{i=1}^n \Indicator(\boldsymbol{\rho}^k_{l,.} = \boldsymbol{y}_{i, k}), \mbox{for}\ \ l=1,\ldots, 2^k. \label{post1}
\end{equation}

If a Dirichlet distribution, i.e. $p(\boldsymbol{w}_{2^k}|\delta_1,\ldots, \delta_{2^k}) = \mbox{Dir}(\boldsymbol{w}_2^k|\delta_1,\ldots, \delta_{2^k}) $ is chosen as a prior distribution over $\boldsymbol{w}_{2^k}$, then it is trivial to see that the posterior distribution is given by
\begin{eqnarray}
p(\boldsymbol{w}_{2^k}|\boldsymbol{y}_{1, k}, \ldots, \boldsymbol{y}_{n, k}) &\propto&p(\boldsymbol{w}_{2^{k}}|\delta_1,\ldots, \delta_{2^k})  \prod_{i=1}^n p(\boldsymbol{Y}_k= \boldsymbol{y}_{i, k}|\boldsymbol{w}_{2^k}), \nonumber \\ 
&\propto& \mbox{Dir}(\boldsymbol{w}_{2^k}| \alpha_1,\ldots, \alpha_{2^k}). \label{post2}
\end{eqnarray}
where $\alpha_l= r_l + \delta_l$, for $l =1, \ldots, 2^k$. In Section \ref{basic_prob} it is shown how to use these weights to obtain meaningful probabilities. 
 
\section{COVID-19 footprint and mortality  }\label{sec:4}

In this section our aim is to study the relationship between comorbidities, symptoms, sex, age and death for all those patients that resulted positive for SARS-CoV-2 virus in a laboratory test in Mexico. Hence, in this section we will use the data of \mtot cases, and for each patient we have the footprint
\begin{eqnarray}
\boldsymbol{y}_{i, 35}&=& (y_{i,1}, y_{i,2}, \ldots, y_{i,35}), \label{data}\\
&=&(\mbox{sex}_i, \mbox{age group}_{i,1:4}, \mbox{comorbidity}_{i,1:9}, \mbox{symptom}_{i,1:19}, \mbox{hospitalized}_i, \mbox{death}_i),  \nonumber \\ 
&=& (s_i, g_{i,1}, \ldots, g_{i,4}, j_{i,1}, \ldots, j_{i,9}, l_{i,1},\ldots, l_{i,19}, h_i, d_i), \nonumber
\end{eqnarray}
for $i=1\ldots, \mtot$. 
\subsection*{Model specification}
To fit the Bernoulli multivariate model we take the footprint $\boldsymbol{y}_{i,35}$ for each patient. Thus, our theoretical sample space is of $2^{35}$ possible outcomes. However, for the four variables of age group there are only four possible results, namely the same person cannot belong to two different age groups at the same time. Hence, the true sample space of our model is of $2^{32}$ outcomes.

The key to handle this huge number of sets is simply to identify the active outcomes in (\ref{post1}), i.e. the unique different outcomes that have been observed in the data. Within the $\mtot$ cases, we have observed $m = \mact$ active outcomes. Setting $k=35$ and denoting $\boldsymbol{\beta}^k_{l,.}$ for $l=1,\ldots, m$ as the active outcomes, we can order the rows of the matrix with all the possible outcomes such that in the first $m$ rows we have the active outcomes, thus  $\boldsymbol{\rho}^k_{l,.} = \boldsymbol{\beta}^k_{l,.}$ for $l=1,\ldots, m$. With this we can write (\ref{post1}) as 
$$r_l = \sum_{i=1}^n \Indicator(\boldsymbol{\rho}^k_{l,.} = \boldsymbol{y}_{i,k}) = \sum_{i=1}^n \Indicator(\boldsymbol{\beta}^k_{l,.} = \boldsymbol{y}_{i,k}),\ 
\mbox{for}\ l=1\ldots, m,$$ and $r_l = 0$, for $l>m$. Then $\alpha_l = r_l + \delta_l$, for $l=1\ldots, m$ and for $l>m$ we have $\alpha_l = \delta_l$. 

To obtain the marginal distribution of $w_l$ along with the distribution of any combination of the weights, we will need to compute 
$$\alpha_0 =  \sum_{l=1}^{2^k} \alpha_l = \sum_{l=1}^{m} r_l + \sum_{l=1}^{2^k} \delta_l.$$
This is the starting point of how to handle a theoretical sample space of $2^{k}$ outcomes, and in the next sections it will become clear that we only need to focus in the $m$ active outcomes. 

\subsubsection*{Prior setting}
Assuming $\delta_l = \delta$ in the Dirichlet prior for the weights, we have $\alpha_0 =   \sum_{l=1}^{m} r_l +  (2^k)(\delta)$. If, for example, $\delta = 1$ is assumed we will be giving uniform weight over the $2^k$ possible outcomes, but this is unrealistic since only $m$ active outcomes out of the $2^k$ possibilities have been observed (with $2^k\gg m$). Also, the sample size $n =\mtot$ cases, is quite large and do not expect drastic changes if we update the information and consider more cases. Hence, taking $\delta =\displaystyle \nu/2^k$, we have 
\begin{equation}
\alpha_0 =   \sum_{l=1}^{m} r_l +  \nu ,\label{a0}
\end{equation}
and giving a small value for $\nu < 1$ our inference will be heavily based on the information available.  

\subsection*{Basic probabilities}\label{basic_prob}
In this section we focus on the weights $w_j$ and $w_{js}$, defined in (\ref{wj}) and (\ref{wjs}) respectively, rather than the complete vector (\ref{wfull}).
The point being the simplified random variables $w_j$ and $w_{js}$ allows us to compute important summaries directly. 

Consider (\ref{data}) and observe that, for example, for any $j$ over the set $\{6,\ldots, 14\}$, the random variable $Y_j$ indicates the presence or absence of comorbidity $j - 5$.  Letting $C_l$ as the event of suffering from comorbidity $l$, with $l\in{1, \ldots, 9}$. Then, from (\ref{py}), (\ref{wj}), (\ref{post2}) and knowing that the marginals of a Dirichlet distribution are Beta random variables, we have 
\begin{eqnarray*}
p(C_{j-5} | \mbox{data}) &{\buildrel d \over =}& p(Y_j= 1|\boldsymbol{y}_{k,1}, \ldots, \boldsymbol{y}_{k,n}),\\
 &{\buildrel d \over =}& w_j |\boldsymbol{y}_{1,k}, \ldots, \boldsymbol{y}_{n,k}   \sim  \mbox{Beta}(w_j|  \eta_j, \alpha_0 - \eta_j),
 \end{eqnarray*}
 where  
 \begin{eqnarray*}
 \eta_j &=& \sum_{l=1}^{2^k} \Indicator(\rho^k_{l,j} = 1) \alpha_l,\\
 &=&  \sum_{l=1}^{m} \Indicator(\beta^k_{l,j} = 1) r_l  +  \delta  \sum_{l=1}^{2^k} \Indicator(\rho^k_{l,j} = 1),\\
 &=&  \sum_{l=1}^{m} \Indicator(\beta^k_{l,j} = 1)  \sum_{i=1}^n \Indicator(\boldsymbol{\beta}^k_{l,.} = \boldsymbol{y}_{i,k})  +  (\delta)  (2^{k-1}),\\
 &=&  \sum_{i=1}^{n} \Indicator(y_{i,j} = 1)  +  \frac{\nu}{2},
\end{eqnarray*}
and $\alpha_0$ is computed as in (\ref{a0}). As a summary measure the expected value can be obtained as
$$\ex\left(p(\mbox{C}_{j-5} | \mbox{data})\right) = \ex(w_j |\boldsymbol{y}_{k,1}, \ldots, \boldsymbol{y}_{k,n})  =  \frac{\eta_j}{\alpha_0}.$$

The same applies to symptoms, hospitalization, death, sex and age group. The complete list of events are 
\begin{itemize}
\item $C_j$ - the patient suffers from comorbidity $j$, with $j\in\{1, \ldots, 9\}$. 
\item $S_l$ - the patient presents symptom $l$, with $l\in\{1, \ldots, 19\}$. 
\item $M$ - the patient is male. Note $M^c = F$, the patient is female. 
\item $A_g$ -  the patient belongs to group age $g$, with $g\in\{1, 2, 3, 4\}$. 
\item $D$ - the patient has died. Note death is the variable 35 in  (\ref{data}). 
\item $H$ - the patient has being hospitalized. Note hospitalization is the variable 34 in  (\ref{data}). 
\end{itemize} 
With this conventions, and assuming $j$, $s$, $l$ and $g$ range over the correct set on each case
\begin{eqnarray*}
p(C_{j-5}, C_{s-5} | \mbox{data}) &{\buildrel d \over =}& w_{j,s} |\boldsymbol{y}_{1, 35}, \ldots, \boldsymbol{y}_{n, 35},\\
p(C_{j-5}, S_{l-14} | \mbox{data}) &{\buildrel d \over =}& w_{j,l} |\boldsymbol{y}_{1, 35}, \ldots, \boldsymbol{y}_{n, 35},\\
p(D| C_{j-5}, \mbox{data}) &{\buildrel d \over =}& \frac{w_{j, 35}}{w_j} |\boldsymbol{y}_{1, 35}, \ldots, \boldsymbol{y}_{n, 35},\\
p(D| C_{j-5}, M, A_{g-1}, \mbox{data}) &{\buildrel d \over =}& \frac{w_{1, g, j, 35}}{w_{1, g, j,}} |\boldsymbol{y}_{1, 35}, \ldots, \boldsymbol{y}_{n, 35},
\end{eqnarray*}
etc., are easy to compute Beta random variables. To see this we can compute distribution of $w_{j,s}$ and $\frac{w_{j,s}}{w_{j}}$ given the data, the remaining expressions are straightforward generalizations. First note that $w_{j,s}$ is defined in expression (\ref{wjs}), hence by the aggregation property of the Dirichlet distribution
 \begin{eqnarray*}
 w_{j,s} |\boldsymbol{y}_{k,1}, \ldots, \boldsymbol{y}_{k,n} &\sim&  \mbox{Beta}(w_{j,s}|  \eta_{j,s}, \alpha_0 - \eta_{j,s}),
\end{eqnarray*}
where 
\begin{eqnarray*}
\eta_{j,s} &=&  \sum_{l=1}^{2^k}  \Indicator(\rho^k_{l,j} = 1)\Indicator(\rho^k_{l,s} = 1) (r_l + \delta), \\
&=& \sum_{i=1}^{n}  \Indicator(y_{i,j} = 1)\Indicator(y_{i,s} = 1) +  \delta \sum_{l=1}^{2^k}  \Indicator(\rho^k_{l,j} = 1) \Indicator(\rho^k_{l,s} = 1),\\
&=& \sum_{i=1}^{n}  \Indicator(y_{i,j} = 1)\Indicator(y_{i,s} = 1) +  (\delta) (2^{k-2}),\\
&=& \sum_{i=1}^{n}  \Indicator(y_{i,j} = 1)\Indicator(y_{i,s} = 1) + \frac{v}{4}.
\end{eqnarray*}
Finally, in Appendix \ref{apB} it is shown that 
$$\frac{w_{j,s}}{w_{j}} |\boldsymbol{y}_{k,1}, \ldots, \boldsymbol{y}_{k,n} \sim \mbox{Beta}\left(\frac{w_{j,s}}{w_{j}}|\eta_{j,s}, \eta_{j}  -\eta_{j,s} \right).$$
and then 
\begin{eqnarray*}
  \ex\left(\frac{w_{j,s}}{w_{j}} |\boldsymbol{y}_{k,1}, \ldots, \boldsymbol{y}_{k,n}\right)  &= & \frac{\eta_{j,s}}{\eta_{j}}.
\end{eqnarray*}
To compute these summaries in the case of females, we need to consider $\Indicator(\beta^k_{l,1} = 0)$ instead of $\Indicator(\beta^k_{l,1} = 1)$, but this is a straightforward adaptation of what has been described.

\subsection*{Prediction}\label{pred}
If a new patient has been diagnosed as SARS-CoV-2 positive, then the information for this patient would be given by its clinical footprint
$$\boldsymbol{y}_{i^*, 35} =(s, g_1, \ldots, g_4, j_1, \ldots, j_9, l_1,\ldots, l_{19}, h, d),$$ where the sex, age group, comorbidities and symptoms are known but the outcome variables hospitalized ($h$) and death ($d$) are unknown. In this  case it is easier to work with the original notation for the weights (\ref{wfull}), since all the variables are needed.

It is straightforward to obtain the predictive probability of death by calculating
\begin{multline*}
p(Y_{35} = 1| (Y_1, \ldots, Y_{33}) = (s, g_1, \ldots, g_4, j_1, \ldots, j_9, l_1,\ldots, l_{19}), \boldsymbol{y}_{1, 35}, \ldots, \boldsymbol{y}_{n, 35}) = \\ \frac{w_{35}(s, g_1, \ldots, g_4, j_1, \ldots, j_9, l_1,\ldots, l_{19}, h=0\ \mbox{or}\ 1, d = 1)}{w_{35}(s, g_1, \ldots, g_4, j_1, \ldots, j_9, l_1,\ldots, l_{19}, h=0\ \mbox{or}\ 1, d=0\ \mbox{or}\ 1)} |\boldsymbol{y}_{1, k}, \ldots, \boldsymbol{y}_{n, k},
\end{multline*}
again is a Beta distribution with parameters $\eta^*$ and $\beta^*$ where 
\begin{eqnarray*}
\alpha_{*} &=& \frac{\nu}{2^{k-1}} + \sum_{l=1}^{m}  \left\{\Indicator(\boldsymbol{\beta}^k_{l,.} = \boldsymbol{y}_{2*,35}) + \Indicator(\boldsymbol{\beta}^k_{l,.} = \boldsymbol{y}_{4*,35})\right\} r_l  \\
\beta_{*} &=&  \frac{\nu}{2^{k-1}} + \sum_{l=1}^{m}  \left\{\Indicator(\boldsymbol{\beta}^k_{l,.} = \boldsymbol{y}_{1*,35}) + \Indicator(\boldsymbol{\beta}^k_{l,.} = \boldsymbol{y}_{3*,35})\right\} r_l .
\end{eqnarray*}
with 
\begin{eqnarray*}
\boldsymbol{y}_{1*,35} &=& (s, g_1, \ldots, g_4, j_1, \ldots, j_9, l_1,\ldots, l_{19}, 0, 0),  \\
\boldsymbol{y}_{2*,35} &=& (s, g_1, \ldots, g_4, j_1, \ldots, j_9, l_1,\ldots, l_{19}, 0, 1),\\
\boldsymbol{y}_{3*,35} &=& (s, g_1, \ldots, g_4, j_1, \ldots, j_9, l_1,\ldots, l_{19}, 1, 0),\\
\boldsymbol{y}_{4*,35} &=& (s, g_1, \ldots, g_4, j_1, \ldots, j_9, l_1,\ldots, l_{19}, 1, 1).
\end{eqnarray*}
Hence, the posterior mean predictive is given by 
$$\ex(p(Y_{35} = 1| (Y_1, \ldots, Y_{33}) = (s, g_1, \ldots, g_4, j_1, \ldots, j_9, l_1,\ldots, l_{19}), \boldsymbol{y}_{1, 35}, \ldots, \boldsymbol{y}_{n, 35}) ) = \frac{\alpha^* }{\alpha^*  + \beta^* },$$
and the predictive is simply the Bernoulli distribution
$$p(z|\alpha_{*}, \beta_{*}) = \left(\frac{\alpha_{*}}{\alpha_{*} + \beta_{*}}\right)^z \left(\frac{\beta_{*}}{\alpha_{*} + \beta_{*}}\right)^{1-z},$$
with $z \in \{0, 1\}$ and $\displaystyle \frac{\alpha_{*}}{\alpha_{*} + \beta_{*}}$ the probability of death.

A couple of comments are in order. First, if prediction about the hospitalization status of the patient is required this can be done in an analogous manner. Second, often only a few subset of variables are needed to make prediction, in this case the variables that are not needed are integrated out as it was done with the hospitalizations. 

\section{Analysis}\label{sec:5}
 Focusing again the analysis on the $n = \mtot$ positive cases, here we compute mortality risk given a particular footprint configuration. Also the expected mortality and correlation between variables is obtained. The section is closed  with a prediction of death outcomes.

\subsection*{Some general summaries}
For patients that have been identified as SARS-CoV-2 positive in a laboratory test in Mexico, it is expected that approximately
\begin{itemize}
\item 10 out of 100 patients to die. 
\item 22 out of 100 patients to be hospitalized.
\item 4 out of 100 patients to be asymptomatic. 
\item 70 out of 100 patients to suffer symptoms such as headache, cough or fever. 
\item 30 out of 100 patients to to suffer difficulty while breathing.
\item 60 out of 100 patients will not have any comorbidity. 
\item 19 out of 100 patients to suffer from hypertension, 16 from obesity and 15 diabetes. 
\end{itemize}
All these summaries were computed via posterior expectations, e.g, $\ex(P(C_j|\mbox{data}))$ or $\ex(P(D|\mbox{data}))$. 

\subsection*{Mortality risk}
\begin{figure}
\begin{center}
			\includegraphics[scale=0.8]{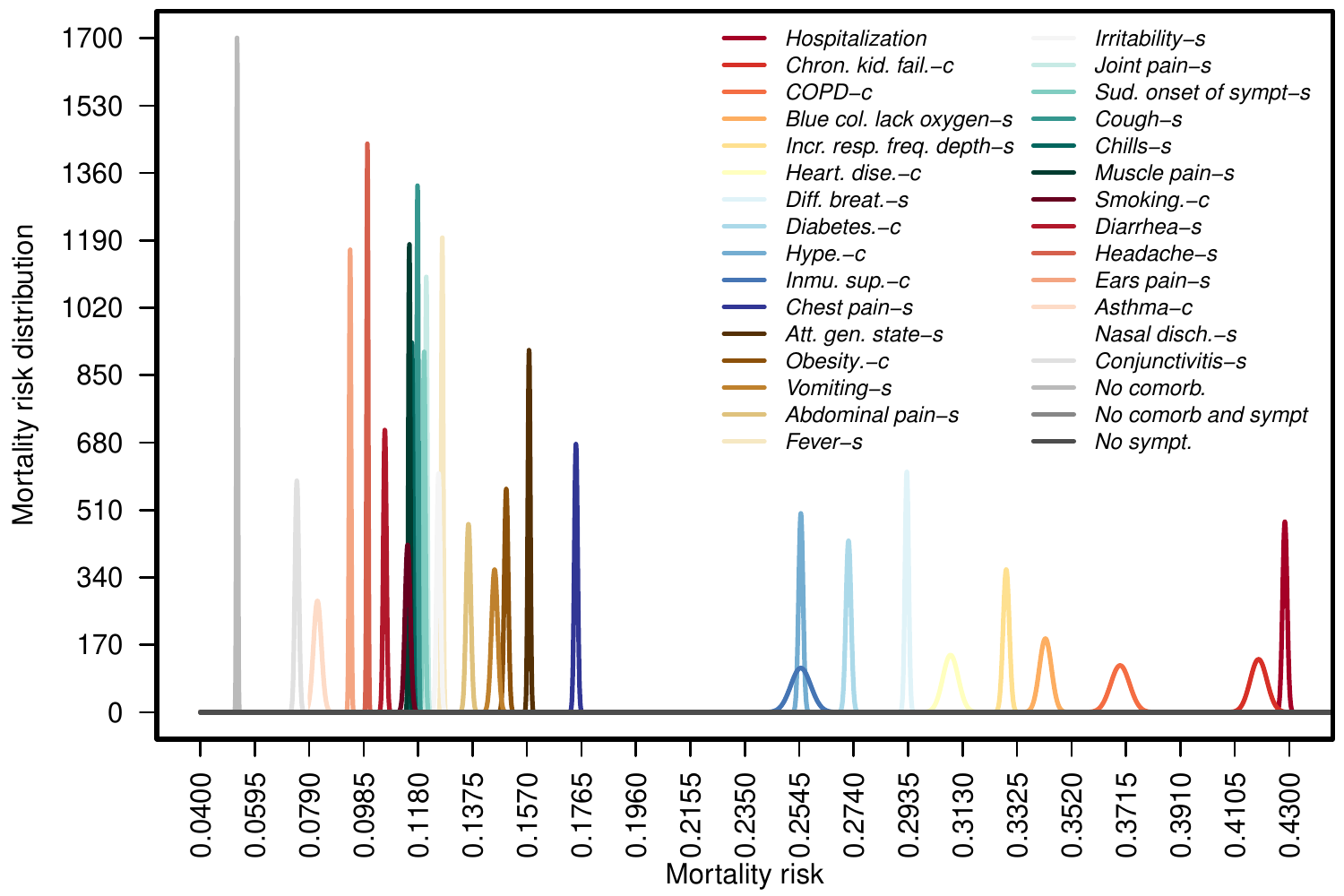}
	\caption{Distribution for the random variables $P(D|C_j, \mbox{data})$, for $j=1\ldots, 9$, $P(D|S_l, \mbox{data})$, for $l=1\ldots, 19$ and $P(D|H, \mbox{data})$.} 
		\label{fig1}
	\vspace{-5pt}
\end{center}
\end{figure}

To identify the comorbidities and symptoms that imply greater mortality risk, we need to obtain the conditional probabilities $P(D|C_j, \mbox{data})$ and $P(D|\mbox{S}_l, \mbox{data})$ that were described in section \ref{basic_prob}. In Figure~\ref{fig1} we display a graphic of these distributions.

First, it is important to note that the colored labels in Figure~\ref{fig1} are ordered following the same order as the conditional distributions over the $x$ axis. Thus, for example, chronic kidney failure, hospitalization, COPD, bluish coloration due to lack of oxygen, etc., are the comorbidities or symptoms that imply greater mortality risk. Second, in these labels we have included a ``-c'' for comorbidity and a ``-s'' for symptom to differentiate each condition. Then, it is easy to identify two groups of comorbidities or symptoms. The first group is formed by the comorbidities or symptoms (including hospitalization) associated with a mortality risk greater than $23\%$, then those with mortality risk lower than $17\%$. The symptoms in the group with higher mortality risk are all related to lungs becoming inflamed, while the comorbidities range from chronic kidney failure, COPD, heart disease diabetes to hypertension. 

It is straightforward to generate this information desegregated by sex and age group. Such cases are displayed in Figure~\ref{fig2}, see also  Table \ref{taba1} in Appendix \ref{tablas}, for more detail. For these cases, we compute, e.g.  $\ex(P(D|C_j, M, A_g, \mbox{data}))$ for the case of comorbidities, males and age group.

\begin{figure}[!htbp]
\begin{center}
			\includegraphics[scale=0.6]{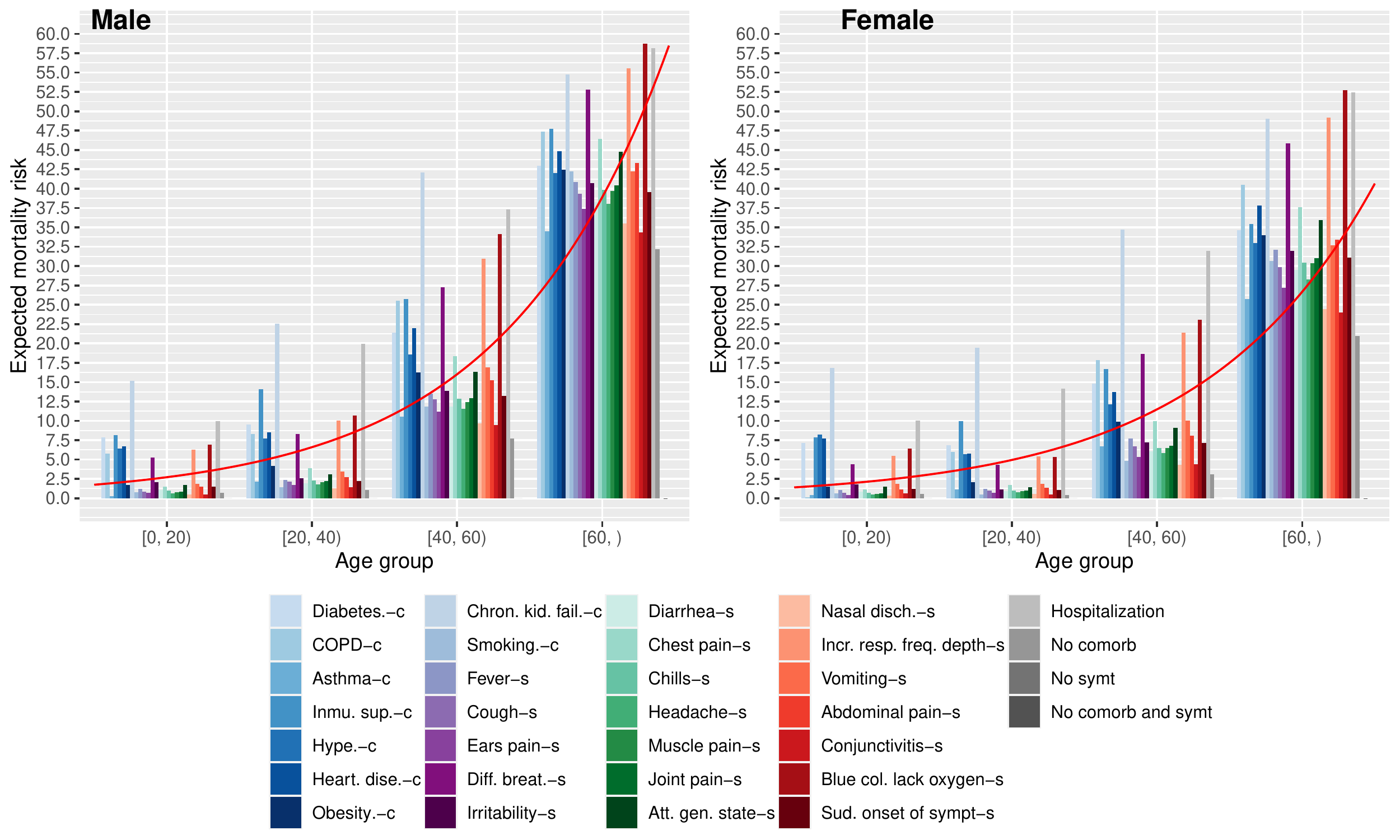}
	\caption{Expected mortality risk given comorbidities/symptoms, sex and age.} 
		\label{fig2}
	\vspace{-5pt}
\end{center}
\end{figure}

If age increases, the mortality risk increases in all cases. This, can clearly be seen via the mortality risk by COVID-19 when the patient has no comorbidities: for patients below $20$ years we can expect $1$ death out of 100 patients (male or female), but increases substantially and for patients over $60$ years old, we expect $30$ deaths out of 100  male patients and $21$ deaths for female patients. This indicates that age is a key variable that increases the mortality risk dramatically even in patients with no comorbidities. Sex is also a key variable. In most cases males have greater mortality risk given comorbidities or symptoms than female patients. Being hospitalized is also associated with a very high mortality risk.

Other interesting findings from Figure~\ref{fig2} (and Table \ref{taba1} in Appendix \ref{tablas}) are the following: 
\begin{itemize}
\item The expected mortality risk (as a percentage) for males and females can be described by a straightforward exponential model of the kind $y=a e^{b x}$ where $y$ is the probability of death and $x$ is the age.  Using the four age groups, and thus assuming  $x = 1, 2, 3$ and $4$. In the case of males, and using least squares fit,  it was obtained that $\hat{a}=1.1$ and  $\hat{b}=0.88$ while for females we have $\hat{a}=0.91$ and  $\hat{b}=0.84$. These are the red lines included in Figure~\ref{fig2}.
\item For cases under 20 years old chronic kidney failure (comorbidity) is associated with the highest mortality risk for males and females, while immunosupression (comorbidity) is the second condition with higher risk for males but for females is hypertension (comorbidity). This difference is interesting since hypertension for males is on $6^{th}$ place. 
\item For cases between 20 and 40 years old chronic kidney failure and immunosuppression are the comorbidities with higher mortality risk for males and females. In third place for males is bluish coloration due to lack of oxygen, while for female patient is diabetes.
\item For cases between 40 and 60 years old chronic kidney failure (comorbidity) is associated with the highest mortality risk for males and females, then for both follows bluish coloration due to lack of oxygen (symptom) and increased respiratory rate and depth (symptom) in second and third places respectively.
\item For cases above 60 years old, bluish coloration due to lack of oxygen (symptom) is associated with the highest mortality risk for males and females, then for males and females follows increased respiratory rate and depth (symptom), and the fourth is chronic kidney failure (comorbidity). 
\end{itemize}

\subsection*{COVID-19 mortality}
Let us assume that $v = 100,000$ people in Mexico were diagnosed as positive for SARS-CoV-2 virus. To appreciate which comorbidities and symptoms have greater impact over COVID-19 mortality, from these $v$ cases, we can approximate the number of people who suffer from comorbidity $C_j$, and this is given by $v p(C_j|\mbox{data})$. Then the approximate the mortality within those with $C_j$ would be $v p(C_j |\mbox{data})  p(D|C_j, \mbox{data})= v p(D, C_j|\mbox{data})$. Thus, to obtain information about the mortality associated with each comorbidity we need the distribution of $p(D, C_j|\mbox{data})$ and we will have a similar expression for the symptoms.

\begin{figure}[!htbp]
\begin{center}
			\includegraphics[scale=.55]{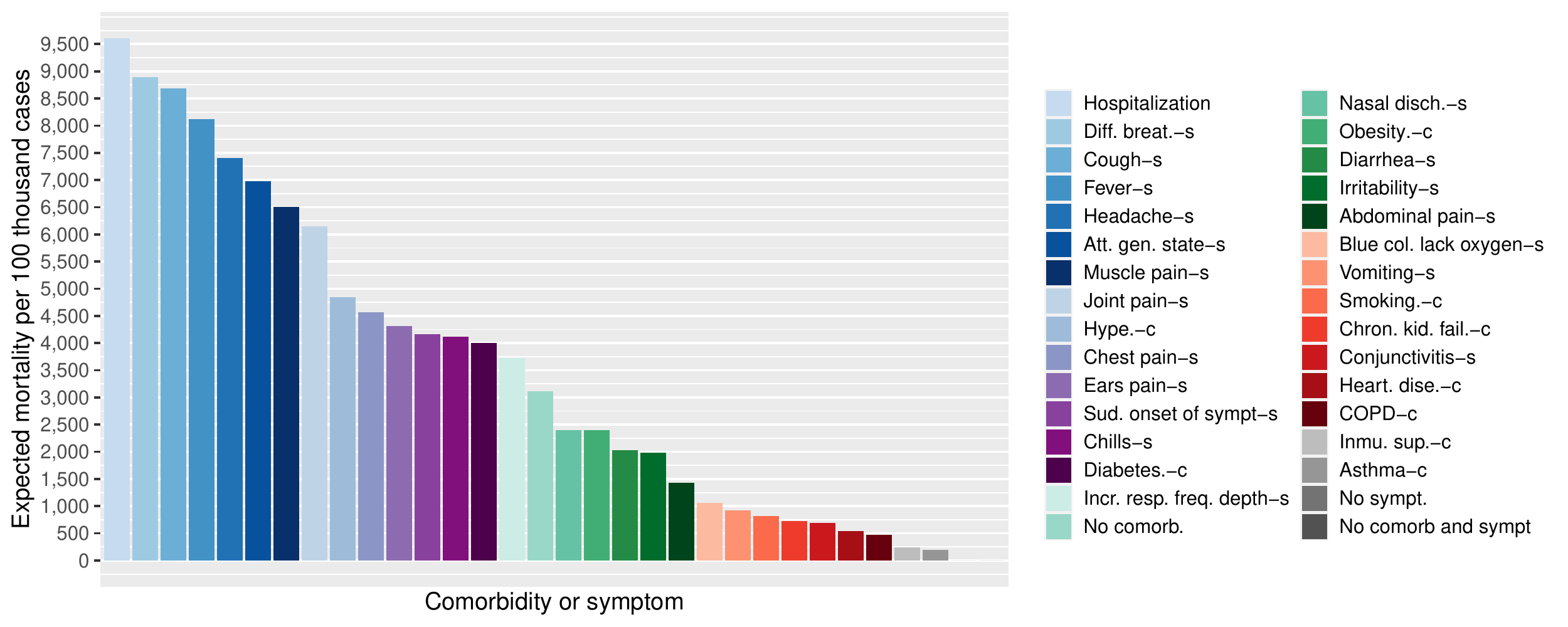}
	\caption{Expected mortality related to each comorbidity or symptom out of 100,000 cases.} 
		\label{fig3}
	\vspace{-5pt}
\end{center}
\end{figure}

Instead of working with with $v p(D, C_j|\mbox{data})$ we obtain its expectation and this is shown in Figure~\ref{fig3} (assuming $v=100,000$ cases).

Hospitalization is clearly associated with the highest mortality, however it is not a comorbidity nor a symptom. Following hospitalization there are seven symptoms, and the one that leads to the a greater mortality is difficulty breathing. The comorbidities that have greater impact over the mortality are hypertension and diabetes, and this is of great concern to health authorities due to the enormous problem of diabetes and hypertension in Mexico. Now we can do the complete exercise to estimate the mortality related to each comorbidity (or symptom) by sex and age group. 

For simplicity, let $Y$ be the distribution of the mortality related to each comorbidity by sex and age group out of $v$ cases (the expression for  symptoms and female cases is similar), then
\begin{eqnarray*}
Y &{\buildrel d \over =}&   v\ p(D, M, A_g, C_j|\mbox{data}),\\
&{\buildrel d \over =}& v\ p(M|\mbox{data}) p(A_g| M,\mbox{data}) p(C_j| A_g, M,\mbox{data}) p(D| C_j, A_g, M,\mbox{data}),\\
&\Rightarrow& \ex(Y)= v\ \ex(p(D, M, A_g, C_j|\mbox{data})).
\end{eqnarray*}  

This is again a constant multiplied by a Beta random variable. The corresponding  expected values are displayed in Figure~\ref{fig4} (see Table \ref{taba3} in Appendix \ref{tablas} for more details).
\begin{figure}[!htbp]
\begin{center}
			\includegraphics[scale=.55]{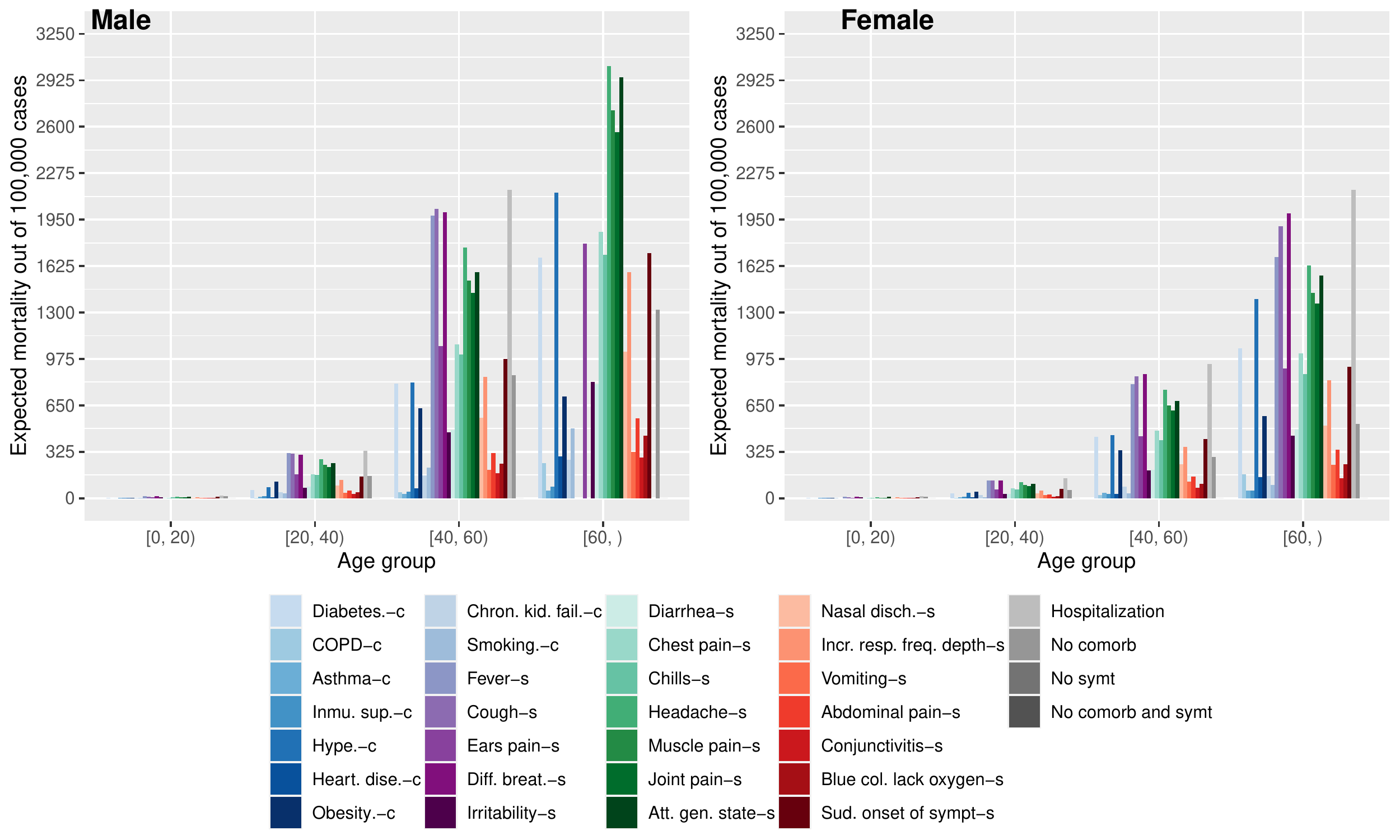}
	\caption{Expected mortality related to each comorbidity (or symptom) by sex and age group out of 100,000 cases.} 
		\label{fig4}
	\vspace{-5pt}
\end{center}
\end{figure}

\subsubsection*{Symptoms, sex and age of the patients}
\begin{figure}[!htbp]
\begin{center}
			\includegraphics[scale=.13]{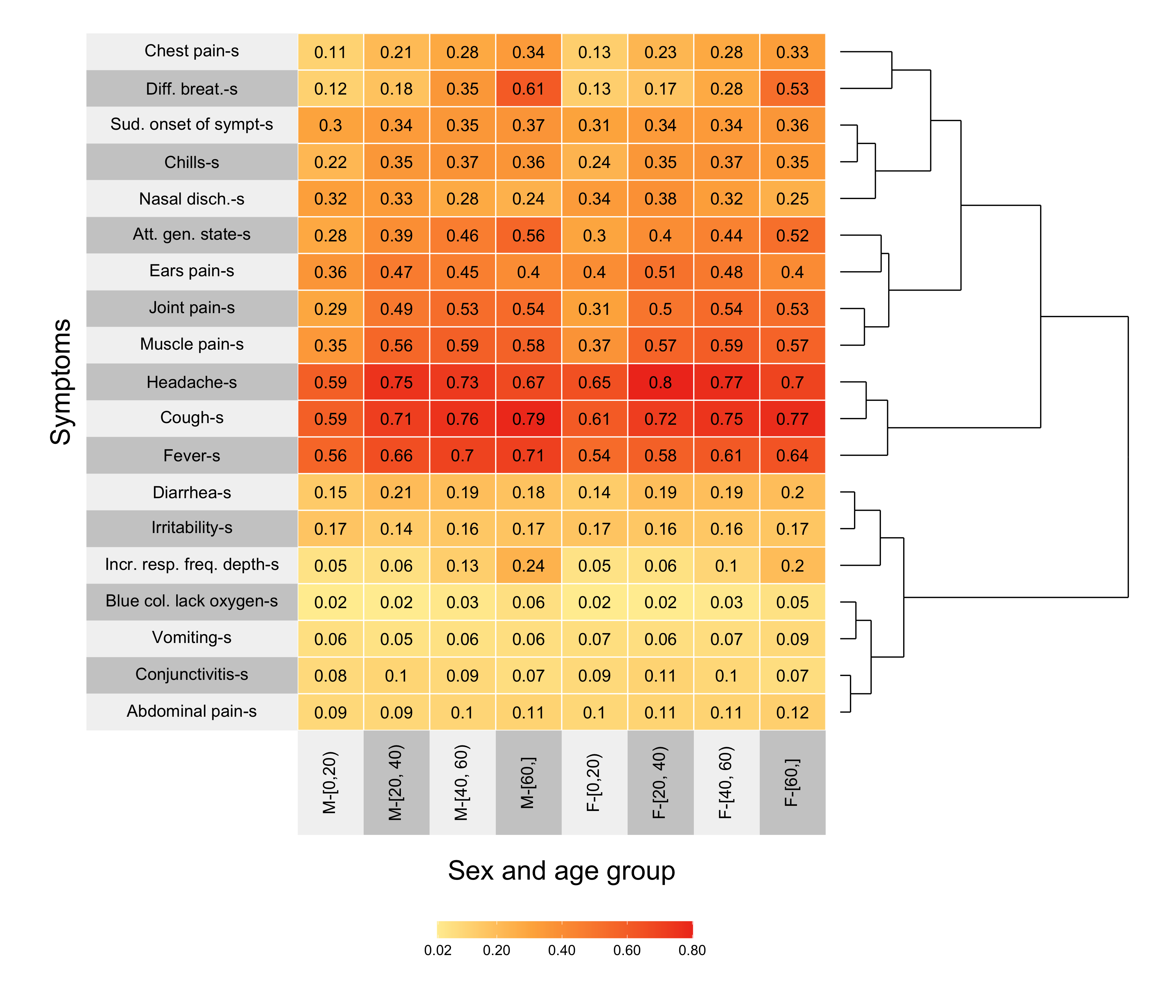}
	\caption{Expected probability of suffering each one of 19 symptoms by sex and age group, for all those who have been diagnosed as SARS-CoV-2 positive in a laboratory test.} 
		\label{heat}
	\vspace{-5pt}
\end{center}
\end{figure}

In the previous section we obtained $p(S_l| M, A_g,\mbox{data})$, $p(S_l| F, A_g,\mbox{data})$ and its expectation. With this it is straightforward to identify if there are patterns of symptoms for all the cases by sex and age group. In Figure~\ref{heat} the expected probability of each symptom given sex and age group is displayed via a heat-map.

Cough, fever and headache are the most common symptoms, but the probability of having the first two increases marginally with age. Also, the mortality risk associated to difficulty in breathing, chest pain and increased respiratory frequency and depth increases with age. This is for both, male and female patients. On the other side, the probability of having nasal discharge decreases with age. 

\subsubsection*{Correlation between comorbidities, symptoms, hospitalization and deaths}
The multivariate Bernoulli model give us the possibility to obtain the correlation between the comorbidities, deaths and hospitalizations. And this is done using equation (\ref{eq3}), but here we need to approximate the posterior mean
$$
\ex\left(\mbox{Corr}(Y_j, Y_s)|\boldsymbol{y}_{k,1}, \ldots, \boldsymbol{y}_{k,n}\right) = \ex\left( \frac{w_{j, s} - w_j w_s}{\sqrt{w_j(1-w_j)}\sqrt{w_s(1-w_s)}}|\boldsymbol{y}_{k,1}, \ldots, \boldsymbol{y}_{k,n} \right),
$$
via classic Monte Carlo, and the idea is described in Appendix \ref{APE}.

\begin{figure}[!htbp]
\begin{center}
			\includegraphics[scale=.55]{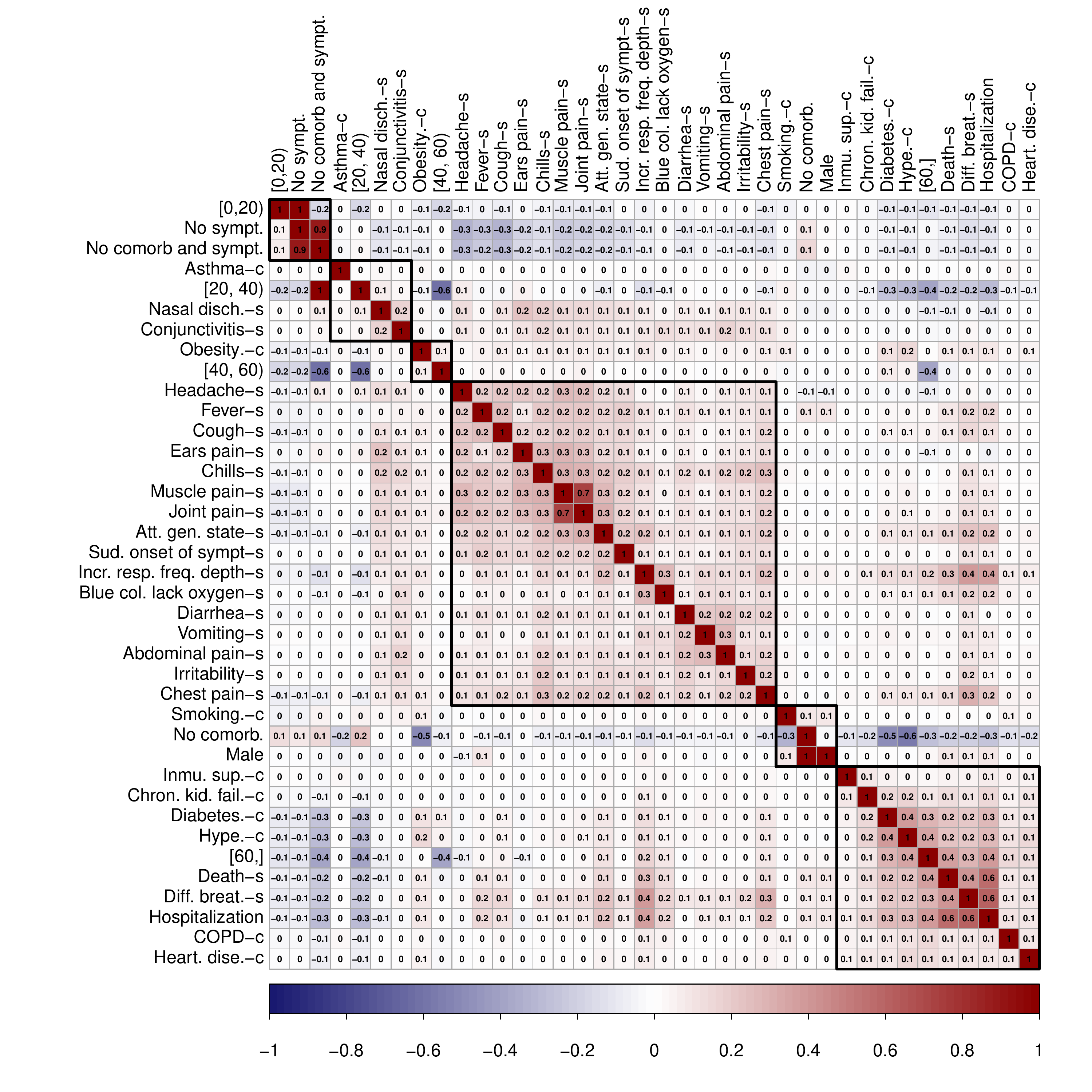}
	\caption{Correlation plot of comorbidities, symptoms, sex, age groups, hospitalization and death.} \label{fig6}
	\vspace{-5pt}
\end{center}
\end{figure}

With the matrix of pairwise correlations a hierarchical clustering is used to obtain groups of variables closely associated to each other. In Figure~\ref{fig6} the correlation plot and the groups are displayed.  

The correlation plot indicates that there are six groups of variables closely related to each other. The group of the outcomes hospitalizations and deaths includes the comorbidities immunosuppression, COPD, heart disease, chronic kidney failure, diabetes and hypertension. The only symptom that is included in this group is difficulty breathing. Finally, the group age of all those greater than 60 years old is also included. Observe that this coincides with Figure~\ref{fig4}, which indicates that larger mortality rates are associated to the variables in this group.

\subsection*{Prediction}
To predict COVID-19 death outcomes using the ideas outlined in Section~\ref{pred}, only six variables are used: sex ($s$), age group ($g$), two comorbidities (diabetes $j_1$ and hypertension $j_2$), one symptom (difficulty breathing $l$) and information about the hospitalizations ($h$).  These variables have been included as the correlation plot in Figure~\ref{fig6}, indicates that these are those with higher correlation with the death outcomes. It could seem that relationships of higher order are not been considered, however we compared this parsimonious model against the complete model with 31 variables and obtained similar results (not shown).

The parameters of the beta distribution are calculated as described in Section \ref{pred}. This can be achieved efficiently concatenating the values of the six variables for the training set, which creates the corresponding footprint for each combination of sex, age group, comorbidities, symptom and hospitalization. These footprints along with the death outcomes are used to generate a contingency table as shown in Table \ref{cont2}. Note that instead of using the footprints 1000, 0100, 0010 and 0001 for age group we only use 1, 2, 3, and 4 and this is as there are only four possible combinations for this variable. 

\begin{table}
\caption{Contingency table for the training set: combination of sex, age group, comorbidities, symptoms, and hospitalizations footprints against death outcomes.}\label{cont2}
\centering
\begin{tabular}{ccc}
$s g j_1 j_2 l h$ & Alive & Death \\ 
010000 & 24,313 & 7\\
010001 & 1,026 & 45 \\ 
\vdots & \vdots        &  \vdots \\
   040111 & 6,314 & 6,896 \\ 
   \vdots &   \vdots        &  \vdots \\
   130011 & 22,378 & 11,164 \\ 
    \vdots & \vdots        &  \vdots \\
   141111 & 7,321 & 10,566 \\
\end{tabular}
\end{table}

Then, the footprints for each combination of of sex, age group, symptoms, comorbidities and hospitalization in the prediction set are obtained. If in the prediction set we have, for example, the footprint 130011, the contingency table in Table \ref{cont2}, can be used to obtain
$$\eta^* = v/2^5 + 11,164\quad \mbox{and}\quad \beta^* =v/2^5  + 22,378,$$
and the posterior mean predictive is given by $\eta^* /(\eta^* +\beta^*) \approx 0.33$. If this is greater than a cut-point $c$ the predicted outcome would be death, otherwise this footprint implies an alive outcome. If a footprint in the prediction set cannot be found in the training set, instead of considering only the prior, sex and age group are used to compute the posterior mean predictive in these cases.

It is important to stress that this model, together  with the ideas outlined in this section, can handle large data bases easily. Indeed, a logistic regression was also implemented to have a benchmark model, and the multivariate Bernoulli model needs less CPU time to calculate the predictions. Note that in the case of the logistic model, age has been included as a continuous variable, as it was used without any categorization.

\subsubsection*{Optimal cut-point} To determine the cut-point $c$ we maximize the proportion of death outputs that are correctly identified (true positive rate or TPR) plus the proportion of alive outcomes that are correctly identified (true negative rate or TNR). Thus, the optimal cut-point maximizes the quantity TPR + TNR. This optimization exercise is performed by randomly dividing the total $\mtot$cases randomly into a training set of $\mtottrain$observations and a prediction set of $100,000$ cases. Then, obtaining the posterior mean predictive for each footprint in the training set and using these to classify the observations with the same footprint in the prediction set. In this last step is where the optimization over $c$ is performed. To avoid any bias from the selection of a particular training set and prediction set, this processes was repeated $100$ times to obtain $\widehat{c}_1, \ldots, \widehat{c}_{100}$, and the optimal cut-point is taken as the mean. The library of R cutpointr was used to find the optimal cut-points, see \cite{cut}. The optimal cut-point for the multivariate Bernoulli and logistic models were $\widehat{c}=0.112$ and $\widehat{c}=0.058$ respectively. The maximum value for TPR + TNR is of $1.777$ and $1.776$ for the multivariate Bernoulli and logistic models respectively.

The process outlined in the preceding paragraph is similar to that of using a  receiver operating characteristic curve (ROC) curve. In this case the quantities (1-TNR,  TPR) are plotted for different values of the cut-point, $c$, and this is the ROC curve. Thus, the optimal $c$ is the one that produces the best possible combination of TNR and TPR. Note however, the values of $c$ are not plotted. It might be seen as a clearer strategy  to have a function to optimize and see explicitly how it changes with the change of the cut-point. However, for completion we mention that the average (over the 100 samples)  area under the ROC curve (AUC) for the multivariate Bernoulli was found to be equal to $0.935$ while for the logistic regression was equal to $0.936$. In general terms classifiers with larger AUC are considered better to discriminate between groups, hence according to this criterion there is virtually no difference between the discrimination ability of both models. That said, under our approach a much wider set of summaries and prediction possibilities are available.

\subsubsection*{20 fold Cross-Validation}
 To test the prediction power of the algorithm we use a 20 fold cross-validation setting: divided the total $\mtot$cases randomly into 20 sets or folds of approximately 79,214 cases each. The first fold is treated as a validation set, and the method is fit on the remaining 19 folds. The confusion matrix is then computed on the observations in the held-out fold, thus generating 19 confusion matrices. Taking the average of these 19 confusion matrices an average confusion matrix $C_1$ was obtained. This procedure is repeated 20 times; each time, a different group of observations is treated as a validation set. This process results in 20 estimates of the confusion matrix, $C_1, C_2, . . . , C_{20}$. The 20-fold Cross-Validation estimate of the overall confusion matrix is computed by averaging these 20 matrices. See Tables \ref{predictive1} and \ref{predictive2}.

\begin{table}
\caption{\label{predictive1} Multivariate Bernoulli: 20-fold cross validation estimate of the confusion matrix (left) and column percentages.}
\centering
{\footnotesize
\begin{tabular}{rrrr}
                    &          & \multicolumn{2}{c}{Actual}\\
\cline{3-4}
                                & Outcomes & Alive & Death \\ 
  \hline
\multirow{2}{*}{Predicted} & Alive & 60,096.9 & 608.8 \\
                                & Death & 10,720.2 & 7,787.9 \\ 
   \hline
\end{tabular}
\quad
\begin{tabular}{rrrr}
                    &          & \multicolumn{2}{c}{Actual}\\
\cline{3-4}
                                & Outcomes & Alive & Death \\ 
  \hline
\multirow{2}{*}{Predicted} & Alive & 84.9\% & 7.3\%\\
                                & Death & 15.1\% & 92.7\% \\ 
   \hline
\end{tabular}}
\end{table}

\begin{table}
\caption{\label{predictive2} Logistic regression: 20-fold cross validation estimate of the confusion matrix (left) and column percentages.}
\centering
{\footnotesize
\begin{tabular}{rrrr}
                    &          & \multicolumn{2}{c}{Actual}\\
\cline{3-4}
                                & Outcomes & Alive & Death \\ 
  \hline
\multirow{2}{*}{Predicted} & Alive & 59,759.3 & 538.0 \\
                                & Death & 11,057.8 & 7,858.8 \\    \hline
\end{tabular}
\quad
\begin{tabular}{rrrr}
                    &          & \multicolumn{2}{c}{Actual}\\
\cline{3-4}
                                & Outcomes & Alive & Death \\ 
  \hline
\multirow{2}{*}{Predicted} & Alive & 84.4\% & 6.4\%\\
                                & Death & 15.6\% & 93.6\% \\ 
   \hline
\end{tabular}}
\end{table}

Hence, via the multivariate Bernoulli we have  (TPR, TNR) = (92.7\%, 84.9\%), while using a logistic regression  (TPR, TNR) = (93.6\%, 84.4\%) is obtained. Classification-wise the results are comparable, and it is worth noting that there is no difference in performance between a model that uses age as a continuous variable and a model that uses age coded as categorical variable.

\section{Discussion}\label{sec:7}
We have used a multivariate Bernoulli distribution to analyze  information about the COVID-19 pandemic in Mexico. In particular we have used the data of patients identified as SARS-CoV-2 positive in a laboratory tests and determined: (1) the comorbidities immunosuppression,  COPD, heart disease, chronic kidney failure, diabetes, hypertension and the symptom difficulty breathing are the variables with stronger association with hospitalizations and deaths; (2) the expected mortality risk for males and females follows an exponential increase which is a function of the age of the patient, and the increase rate of the curve for males is steeper than for female patients; (3) the comorbidities associated with greater expected mortality in Mexico are hypertension, diabetes and obesity, and this is mainly because the prevalence of these chronic diseases in Mexico is very high; (4) cough, fever and headache are the most common symptoms, but the probability of having the first two increases marginally with age. Also, the probability of suffer difficulty breathing, chest pain and increased respiratory frequency and depth increases with age; (5) using only two comorbidities (diabetes and hypertension), one symptom (difficulty breathing), information about hospitalizations, the age group and sex of the patients it is possible to predict the mortality of the COVID-19 disease with the following accuracy: 92.7\% of death outputs have been correctly classified and  84.9\% of alive outputs have been correctly identified.

Finally, it is important to mention that via the logistic regression we tried to select a small subset of variables that could describe death outcomes effectively (out of the 31 variables).  We used the library of R bigstep (\cite{bigstep}) and then try all the available criteria (AIC, BIC and modifications), and were only able to reduce to 16 variables (this analysis is not included in the paper). Via the multivariate Bernoulli and obtaining the correlation matrix is easy to see that there are six variables closely correlated to the death outcomes.


\section*{Acknowledgements} 
The first and second authors are grateful for the support of PAPIIT-UNAM projects IA103220, IV100220 and IG100221. 


\clearpage

\begin{appendix}

\section{Tables}\label{tablas}
These tables have been computed using the information for all those patients who tested positive for SARS-CoV-2 in a laboratory test. 

\begin{table}
\caption{\label{taba1} $\ex\left(P(D| S_l, A_g,  Sex)\right)$ and $\ex\left(P(D| C_j, A_g, \mbox{Sex})\right)$ in percentages. }
\centering
{\scriptsize
\begin{tabular}{lcccccccc}
  \hline
  Sex & \multicolumn{4}{c}{Male} & \multicolumn{4}{c}{Female}\\
  \hline
Com. or Sympt.\textbackslash{} Age group & $< 20 $ & $[20, 40)$ & $[40, 60)$ & $\geq 60$ & $< 20$ & [20, 40) & [40, 60) & $\geq 60$ \\ 
  \hline
 Diabetes.-c & 7.9 & 9.5 & 21.4 & 42.9 & 7.1 & 6.9 & 14.8 & 34.7 \\ 
  COPD-c & 5.8 & 8.3 & 25.6 & 47.4 & 0.1 & 6.0 & 17.9 & 40.5 \\ 
  Asthma-c & 0.3 & 2.1 & 10.6 & 34.5 & 0.4 & 1.1 & 6.7 & 25.8 \\ 
  Inmu. sup.-c & 8.2 & 14.1 & 25.8 & 47.7 & 7.8 & 10.0 & 16.7 & 35.4 \\ 
  Hype.-c & 6.4 & 7.7 & 18.6 & 42.0 & 8.2 & 5.7 & 12.2 & 33.0 \\ 
  Heart. dise.-c & 6.7 & 8.5 & 22.0 & 44.8 & 7.7 & 5.7 & 13.7 & 37.8 \\ 
  Obesity.-c & 1.7 & 4.2 & 16.3 & 42.4 & 1.5 & 2.1 & 9.9 & 34.0 \\ 
  Chron. kid. fail.-c & 15.1 & 22.5 & 42.1 & 54.7 & 16.8 & 19.5 & 34.7 & 49.0 \\ 
  Smoking.-c & 0.8 & 1.5 & 11.9 & 42.2 & 0.6 & 0.5 & 4.9 & 30.7 \\ 
  Fever-s & 1.2 & 2.4 & 13.5 & 40.9 & 1.1 & 1.2 & 7.7 & 32.1 \\ 
  Cough-s & 0.9 & 2.1 & 12.8 & 39.3 & 0.7 & 1.0 & 6.7 & 29.9 \\ 
  Ears pain-s & 0.7 & 1.7 & 11.2 & 37.4 & 0.4 & 0.7 & 5.3 & 27.2 \\ 
  Diff. breat.-s & 5.3 & 8.3 & 27.3 & 52.8 & 4.4 & 4.3 & 18.6 & 45.9 \\ 
  Irritability-s & 2.1 & 2.6 & 13.9 & 40.7 & 1.8 & 1.1 & 7.2 & 32.0 \\ 
  Diarrhea-s & 1.0 & 2.0 & 11.8 & 38.2 & 0.8 & 0.9 & 6.1 & 29.5 \\ 
  Chest pain-s & 1.5 & 3.9 & 18.3 & 46.4 & 1.2 & 1.7 & 9.9 & 37.6 \\ 
  Chills-s & 1.0 & 2.3 & 12.9 & 39.9 & 0.7 & 1.0 & 6.5 & 30.4 \\ 
  Headache-s & 0.6 & 1.8 & 11.5 & 38.1 & 0.5 & 0.8 & 5.8 & 28.3 \\ 
  Muscle pain-s & 0.8 & 2.1 & 12.5 & 39.7 & 0.6 & 0.9 & 6.5 & 30.4 \\ 
  Joint pain-s & 0.9 & 2.2 & 12.9 & 40.4 & 0.6 & 1.0 & 6.8 & 31.0 \\ 
  Att. gen. state-s & 1.7 & 3.1 & 16.4 & 44.8 & 1.5 & 1.5 & 9.1 & 35.9 \\ 
  Nasal disch.-s & 0.5 & 1.3 & 9.7 & 35.5 & 0.4 & 0.5 & 4.3 & 24.5 \\ 
  Incr. resp. freq. depth-s & 6.2 & 10.1 & 31.0 & 55.5 & 5.5 & 5.4 & 21.4 & 49.2 \\ 
  Vomiting-s & 1.8 & 3.4 & 16.9 & 42.2 & 1.9 & 1.8 & 10.0 & 32.6 \\ 
  Abdominal pain-s & 1.5 & 2.7 & 15.2 & 43.3 & 1.1 & 1.4 & 8.1 & 33.4 \\ 
  Conjunctivitis-s & 0.5 & 1.5 & 9.4 & 34.4 & 0.6 & 0.5 & 4.4 & 24.0 \\ 
  Blue col. lack oxygen-s & 6.9 & 10.7 & 34.2 & 58.7 & 6.4 & 5.3 & 23.1 & 52.7 \\ 
  Sud. onset of sympt-s & 1.5 & 2.2 & 13.2 & 39.5 & 1.2 & 1.1 & 7.1 & 31.1 \\ 
  Hospitalization & 9.9 & 19.9 & 37.3 & 58.2 & 10.1 & 14.1 & 31.9 & 52.5 \\ 
  No comorb & 0.7 & 1.1 & 7.7 & 32.2 & 0.5 & 0.4 & 3.1 & 21.0 \\ 
  No symt & 0.0 & 0.0 & 0.0 & 0.0 & 0.0 & 0.0 & 0.0 & 0.0 \\ 
  No comorb and symt & 0.0 & 0.0 & 0.0 & 0.0 & 0.0 & 0.0 & 0.0 & 0.0 \\ 
   \hline
\end{tabular}}
\end{table}

\begin{table}
\caption{\label{taba2}$\ex\left(P(\mbox{Sex})\right) $ top of the table and $\ex\left(P(A_g |\mbox{Sex})\right) $ in percentages. }
\centering
{\footnotesize
\begin{tabular}{rrr}
  \hline
Age group\textbackslash{}Sex & Male (56) & Female (44) \\ 
  \hline
$<20$ & 4.2 & 4.4 \\ 
  $[20, 40)$  & 36.7 & 39.3 \\ 
  $[40, 60)$ & 37.8 & 37.8 \\ 
  $\geq 60$ & 21.3 & 18.5 \\ 
   \hline
\end{tabular}}
\end{table}

\begin{table}
\caption{\label{taba3} $\ex\left(P(D|A_g, Sex)\right)$, $\ex\left(P(H|A_g, Sex)\right)$, $\ex\left(P(\mbox{C}_j|A, Sex)\right)$ and $\ex\left(P(\mbox{S}_l |A_g, Sex)\right)$ in percentages. Estimates for $Sex=M$ and $F$ have been included}
\centering
{\scriptsize
\begin{tabular}{lcccccccc}
  \hline
  Sex & \multicolumn{4}{c}{Male} & \multicolumn{4}{c}{Female}\\
  \hline
 Com. or Sympt.\textbackslash{} Age group & $< 20 $ & $[20, 40)$ & $[40, 60)$ & $\geq 60$ & $< 20$ & [20, 40) & [40, 60) & $\geq 60$ \\ 
  \hline
 Diabetes.-c & 0.8 & 3.0 & 17.9 & 33.2 & 0.9 & 2.9 & 17.1 & 36.6 \\ 
  COPD-c & 0.1 & 0.2 & 0.8 & 4.4 & 0.1 & 0.2 & 0.8 & 4.9 \\ 
  Asthma-c & 3.6 & 2.1 & 1.5 & 1.3 & 3.1 & 2.9 & 3.2 & 2.5 \\ 
  Inmu. sup.-c & 1.2 & 0.5 & 0.8 & 1.5 & 0.9 & 0.5 & 1.1 & 1.9 \\ 
  Hype.-c & 0.7 & 4.9 & 20.8 & 43.1 & 0.6 & 3.6 & 21.5 & 51.1 \\ 
  Heart. dise.-c & 0.7 & 0.5 & 1.5 & 5.6 & 0.6 & 0.5 & 1.2 & 4.8 \\ 
  Obesity.-c & 4.9 & 13.9 & 18.4 & 14.2 & 4.6 & 13.1 & 20.1 & 20.4 \\ 
  Chron. kid. fail.-c & 0.5 & 0.9 & 1.8 & 4.2 & 0.3 & 0.6 & 1.4 & 3.8 \\ 
  Smoking.-c & 3.0 & 11.3 & 8.6 & 9.8 & 1.4 & 5.9 & 4.3 & 3.7 \\ 
  Fever-s & 56.5 & 65.8 & 70.0 & 70.5 & 53.6 & 58.4 & 61.1 & 63.6 \\ 
  Cough-s & 59.2 & 71.4 & 75.5 & 78.9 & 60.8 & 71.7 & 75.0 & 77.0 \\ 
  Ears pain-s & 36.4 & 47.4 & 45.3 & 40.4 & 39.6 & 51.0 & 48.4 & 40.4 \\ 
  Diff. breat.-s & 11.9 & 18.0 & 35.1 & 60.5 & 12.8 & 16.7 & 27.6 & 52.6 \\ 
  Irritability-s & 16.6 & 14.4 & 15.8 & 16.9 & 16.9 & 15.8 & 16.0 & 16.5 \\ 
  Diarrhea-s & 14.6 & 20.5 & 19.3 & 17.7 & 14.1 & 19.2 & 19.2 & 19.6 \\ 
  Chest pain-s & 11.3 & 21.3 & 28.1 & 34.1 & 12.7 & 22.8 & 28.2 & 32.6 \\ 
  Chills-s & 22.2 & 35.1 & 37.5 & 36.2 & 23.5 & 35.4 & 37.1 & 34.6 \\ 
  Headache-s & 59.1 & 75.1 & 72.8 & 67.3 & 65.0 & 80.1 & 77.2 & 69.6 \\ 
  Muscle pain-s & 34.9 & 55.7 & 58.5 & 57.9 & 37.1 & 56.8 & 59.3 & 57.2 \\ 
  Joint pain-s & 29.0 & 48.7 & 53.2 & 53.7 & 31.1 & 49.7 & 53.9 & 53.1 \\ 
  Att. gen. state-s & 28.4 & 39.4 & 46.3 & 55.8 & 29.6 & 39.7 & 44.3 & 52.5 \\ 
  Nasal disch.-s & 31.7 & 33.5 & 27.7 & 24.4 & 34.0 & 38.3 & 32.4 & 25.2 \\ 
  Incr. resp. freq. depth-s & 5.0 & 6.3 & 13.1 & 24.2 & 5.0 & 5.8 & 9.9 & 20.3 \\ 
  Vomiting-s & 6.4 & 5.3 & 5.7 & 6.5 & 6.8 & 6.4 & 6.9 & 8.6 \\ 
  Abdominal pain-s & 9.0 & 9.3 & 10.0 & 11.0 & 9.8 & 10.7 & 11.0 & 12.3 \\ 
  Conjunctivitis-s & 8.3 & 10.2 & 8.9 & 7.0 & 8.8 & 11.1 & 9.7 & 7.1 \\ 
  Blue col. lack oxygen-s & 2.0 & 1.8 & 3.4 & 6.3 & 2.0 & 1.8 & 2.6 & 5.4 \\ 
  Sud. onset of sympt-s & 30.2 & 33.9 & 35.4 & 36.8 & 30.7 & 34.1 & 34.5 & 35.7 \\ 
  Death-s & 0.9 & 1.9 & 11.5 & 37.9 & 0.8 & 0.9 & 6.1 & 28.6 \\ 
  Hospitalization & 8.5 & 8.2 & 27.7 & 59.3 & 7.2 & 5.6 & 17.4 & 49.8 \\ 
  No comorb & 86.5 & 70.9 & 53.4 & 34.7 & 88.8 & 75.8 & 54.6 & 29.9 \\ 
  No symt & 9.3 & 4.6 & 3.6 & 2.2 & 8.4 & 3.7 & 3.4 & 2.8 \\ 
  No comorb and symt & 8.5 & 3.9 & 2.7 & 1.3 & 7.9 & 3.2 & 2.5 & 1.6 \\ 
   \hline
 \end{tabular}}
\end{table}

\section{Distributional results}\label{apB}
Let 
\begin{eqnarray*}
w_{j,-s} &=& \sum_{l=1}^{2^k}  \Indicator{(\rho^k_{l,j} =1)}  \Indicator{(\rho^k_{l,s} \neq 1)}\ w_k(\boldsymbol{\rho}^k_{l,.}),\\
\eta_{j,-s} &=&  \sum_{l=1}^{2^k} \left( \Indicator(\rho^k_{l,j} = 1) \Indicator(\rho^k_{l,s} \neq 1)\right) (r_l + \delta_l)
\end{eqnarray*}
with equivalent definitions for $w_{s,-j} $ and $\eta_{s,-j}$.
First, it is easy to see that
\begin{eqnarray*}
 w_{j,s} + w_{j,-s} &=& \sum_{l=1}^{2^k}  (\Indicator{(\rho^k_{l,j} =1)}  \Indicator{(\rho^k_{l,s} = 1)} +\Indicator{(\rho^k_{l,j} =1)}  \Indicator{(\rho^k_{l,s} \neq1)})\ w_k(\boldsymbol{\rho}^k_{l,.}), \\
&=&  \sum_{l=1}^{2^k}  \left(  \Indicator{(\rho^k_{l,s} = 1)} + \Indicator{(\rho^k_{l,s} \neq1)} \right) \Indicator{(\rho^k_{l,j} =1)} w_k(\boldsymbol{\rho}^k_{l,.}) ,\\ 
&=&  \sum_{l=1}^{2^k} \Indicator{(\rho^k_{l,j} =1)} w_k(\boldsymbol{\rho}^k_{l,.}) = w_{j},
\end{eqnarray*}
and in the same manner $\eta_{j,s} + \eta_{j,-s} = \eta_j$.

\subsubsection*{Distribution of $w_{j,s}/{w_j}$}
We have 
$$p(w_{j,s}, w_{j,-s}| \eta_{j,s}, \eta_{j,-s}, \alpha_0)=\mbox{Dir}(w_{j,s}, w_{j,-s}|\eta_{j,s}, \eta_{j,-s}, \alpha_0  - \eta_{j,s} -  \eta_{j,-s}).$$

Performing  the transformation $U=w_{j,s}$ and $V = w_{j,s} + w_{j,-s}$ (observe that $V = w_j$) the Jacobian is equal to $1$ and the density is given by
\begin{eqnarray*}
p(u, v) &=& p_{w_{j,s}, w_{j,-s}}(u, v-u| \eta_{j,s}, \eta_{j,-s}, \alpha_0), \\
&\propto& u^{\eta_{j,s} - 1} (v-u)^{\eta_{j,-s} - 1} (1-v)^{\alpha_0 - \eta_{j,s} - \eta_{j,-s} }.
\end{eqnarray*}
Here $u > 0$, $v > u$ and $1 > v$, thus this is a valid probability distribution function. 

Now transforming $X=V$ and $Y=\frac{U}{V}$, then the Jacobian is equal to $x$ and the density is given by
\begin{eqnarray*}
p(x, y) &=& x\ p_{U, V}(x y, x| \eta_{j,s}, \eta_{j,-s}, \alpha_0), \\
&\propto& x (x y)^{\eta_{j,s} - 1} (x -xy)^{\eta_{j,-s} - 1} (1-x)^{\alpha_0 - \eta_{j,s} - \eta_{j,-s} }\\
&\propto&  x^{\eta_{j,s} +\eta_{j,-s} - 1}  (1-x)^{\alpha_0 - \eta_{j,s} - \eta_{j,-s} }\ y^{\eta_{j,s} - 1} (1-y)^{\eta_{j,-s} - 1}
\end{eqnarray*}
Then $X \perp Y$, where $X\sim \mbox{Beta}(x|\eta_j, \alpha_0 - \eta_j)$ and $Y \sim \mbox{Beta}(y|\eta_{j,s}, \eta_j - \eta_{j,s})$. Note that $\eta_j  = \eta_{j,-s} + \eta_{j,s}$.

Since $Y = \frac{w_{j,s}}{w_j}$, then 
\begin{eqnarray*}
\frac{w_{j,s}}{w_j} &\sim& \mbox{Beta}\left(\frac{w_{j,s}}{w_j}|\eta_{j,s}, \eta_j - \eta_{j,s}\right),\\
&\Rightarrow& \ex\left(\frac{w_{j,s}}{w_j} \right) = \frac{\eta_{j,s}}{\eta_j}.
\end{eqnarray*}

\section{Posterior expectation of the correlation}\label{APE}
We want to approximate
\begin{equation*}
\ex\left(\mbox{Corr}(Y_j, Y_s)|\boldsymbol{y}_{k,1}, \ldots, \boldsymbol{y}_{k,n}\right) = \ex\left( \frac{w_{j, s} - w_j w_s}{\sqrt{w_j(1-w_j)}\sqrt{w_s(1-w_s)}}|\boldsymbol{y}_{k,1}, \ldots, \boldsymbol{y}_{k,n} \right).
\end{equation*}
First, note that it is straightforward to generate samples from
\begin{eqnarray*}
 (w_{j,s}, w_{j,-s}, w_{s,-j}) &\sim&  \mbox{Dir}(w_{j,s}, w_{j,-s}, w_{s,-j}|\eta_{j,s}, \eta_{j,-s}, \eta_{s, -j}, \alpha_0  - \eta_{j,s} -  \eta_{j,-s}-  \eta_{s,-j}),\\
 &\sim&  \mbox{Dir}(w_{j,s}, w_{j,-s}, w_{s,-j}|\eta_{j,s}, \eta_{j} - \eta_{j,s}, \eta_{s} - \eta_{j,s}, \alpha_0 + \eta_{j,s} -  \eta_{j} -  \eta_{s} ),
 \end{eqnarray*}
and with each sample we can compute $w_j = w_{j,s} + w_{j,-s}$, $w_s = w_{j,s} + w_{s,-j}$ and then 
$$\xi=g(w_{j,s}, w_{j}, w_{s}) = \frac{w_{j, s} - w_j w_s}{\sqrt{w_j(1-w_j)}\sqrt{w_s(1-w_s)}}.$$
Thus, generating $N$ samples and computing $\xi_1, \ldots,\xi_N$ we can approximate
$$\ex\left(\mbox{Corr}(Y_j, Y_s)|\boldsymbol{y}_{k,1}, \ldots, \boldsymbol{y}_{k,n}\right)  \approx \frac{1}{N}\sum_{l=1}^N \xi_l,$$
and this is a usual Monte Carlo approximation.
\end{appendix}

\end{document}